\documentclass{pas}

\usepackage{multirow}
\usepackage[nopatch]{microtype}
\usepackage{booktabs}
\usepackage{natbib}
\usepackage{threeparttable}

\makeatletter
\providecommand{\@thehead}{}    
\providecommand{\first@page}{1} 
\providecommand{\last@page}{\pageref{LastPage}} 
\makeatother

\begin{document}

\lefttitle{Publications of the Astronomical Society of Australia}
\righttitle{Cambridge Author}



\title{Chemical Enrichment of Isolated Relic Galaxy Mrk1216}

\author{\sn{Erdim} \gn{M. Kıyami}$^{1,2}$, \sn{Gülmez} \gn{Emine}$^{1,2}$ and \sn{Hüdaverdi} \gn{Murat}$^{1}$}

\affil{$^1$Department of Physics, Yıldız Technical University, Istanbul, 34220, Turkey}
\affil{$^2$Graduate School of Natural and Applied Sciences, Yıldız Technical University, Istanbul 34220, Turkey}

\corresp{M.K. Erdim, Email: mkiyami@yildiz.edu.tr}


\history{(Received xx xx 2024; revised xx xx 2025; accepted xx xx 2025)}

\begin{abstract}
In this study, we investigate the chemical enrichment and structural evolution of the isolated elliptical relic galaxy Mrk1216 through X-ray observations. As a red-nugget relic, Mrk1216 provides a rare window into the early Universe, owing to its minimal interaction with the surrounding environment. Using data from the \textit{XMM-Newton} telescope, we model the X-ray emission of its interstellar medium to derive radial temperature and abundance profiles. We find that the central region exhibits an elevated [Mg/Fe] ratio compared to typical early-type galaxies, consistent with a brief but intense star formation episode during its early assembly—a hallmark of relic systems. The nearly flat SNIa ratio profile ($R_{Ia} \sim 0.17$) extending to $\sim0.42R_{500}$ supports an early-enrichment scenario. These results highlight the importance of relic galaxies as benchmarks for studying early galaxy evolution and chemical enrichment. Future high-resolution missions and more advanced theoretical models incorporating more realistic initial mass functions are needed to fully assess their implications.

\vspace{1em}
\noindent\textbf{Note:} Accepted for publication in \textit{Publications of the Astronomical Society of Australia (PASA)}.
\end{abstract}


\maketitle

\section{Introduction} \label{introduction}

The $\Lambda$CDM model posits that structures in the Universe formed hierarchically—smaller structures emerged first and later merged into larger ones. This scenario is commonly referred to as a bottom-up scenario in the literature. Massive elliptical galaxies constitute one specific example within this broader model. Specifically, the hierarchical development of high-mass elliptical galaxies at $z = 0$ occurred in two stages. First, gas-rich (wet) mergers formed the galaxy’s core and facilitated the creation of its central supermassive black hole (SMBH). The dense gas content of these wet mergers triggered an intense phase of star formation. This rapid, dissipative process was largely completed by $z \sim 2$. The compact galaxies formed during this stage are referred to as red nuggets \citep{oser2010two, ferre2017two, buote2018luminous, werner2018digging, buote2019extremely, schnorr2021puzzling}.

A slower, long-term process followed the first stage. In this second stage, red-nuggets began to merge with adjacent lower-mass systems. However, since much of the dense gas in those systems had already been converted into stellar mass during the initial stage, the gas ratio was significantly reduced. As a result, these mergers are classified as dry-mergers that do not trigger star formation and are relatively collisionless. Although their impact on the centre of the galaxy is generally negligible, they do contribute to its overall size and stellar mass. This slow accretion phase continued until $z = 0$. This two-stage evolution is supported by many observations, semi-analytic models and simulations \citep[and references therein]{spolaor2010early, oser2010two, ferre2017two, buote2018luminous, werner2018digging, buote2019extremely}.

Nevertheless, the stochastic nature of mergers prevents some galaxies from progressing to the dry-merger phase, leading to an interrupted evolutionary pathway \citep{quilis2013expected, werner2018digging}. As a result, these galaxies bypass the second stage and have largely preserved the properties of red nuggets formed around $z \sim 2$.

NGC 1277 is the first confirmed relic at low redshift \citep{trujillo2013ngc}. It shows high rotational velocity, large central velocity dispersion, and an over-massive black hole at its centre—features that reveal its relic nature. Subsequently, PGC032873 and Mrk1216 have also been identified as relics based on similar structural and dynamical properties. Analyses of their stellar populations and star-formation histories further support their classification as red-nugget-like relics \citep{ferre2017two, walsh2017black}. Notably, \citet{ferre2017two} also introduced the concept of a "degree of relic" to describe variations within this class of galaxies.

Consequently, these compact systems with old stellar populations ($\gtrsim 10$ Gyr) are considered relics that have undergone little structural evolution since formation \citep{ferre2015massive}. Extensive studies have modelled their mass, metallicity, velocity fields, luminosity, size, stellar populations, and dark matter fractions, highlighting clear distinctions between red-nugget-like relics, other compact galaxies, and present-day massive ellipticals \citep{yildirim2015mrk, yildirim2017structural, werner2018digging, buote2019extremely}.

Building on these findings, the INSPIRE (INvestigating Stellar Population In RElics) project conducted the most comprehensive, systematic study to date on relic galaxies between $0.1<z<0.4$ \citep{spiniello2021inspire_a}. The project introduced a method to identify relic galaxies, quantified their degree of relicness (DoR), and measured their spectral and morphological properties across a broad wavelength range (NIR–VIS–UVB) \citep{spiniello2021inspire_b, d2023inspire, martin2023inspire, spiniello2024inspire, maksymowicz2024inspire, scognamiglio2024inspire}. Among the main results, relics were found to have higher stellar velocity dispersion than non-relic galaxies of the same stellar mass and to require a bottom-heavy IMF with an enhanced dwarf-to-giant ratio \citep{maksymowicz2024inspire}. The project also showed that [Mg/Fe] and metallicity correlate with relicness \citep{spiniello2024inspire} and confirmed that relics can exist both in dense environments, such as clusters, and in isolation \citep{scognamiglio2024inspire}.

\citet{lisiecki2023first} identified 77 relics at intermediate redshifts ($0.5<z<1.0$) using the VIPERS survey. Expanding on this work, \citet{siudek2023environments} investigated the physical properties and environmental densities of 42 relic galaxies, confirming their diversity of environments.

While these studies provided extensive insights into the stellar and structural properties of relic galaxies, they lacked coverage of their X-ray characteristics. Yet, observations of the hot, X-ray-emitting interstellar medium (ISM) are also crucial for understanding the nature of relics. Since observing the ISM of red nuggets at $z \sim 2$ remains exceedingly challenging with current X-ray instrumentation due to their distance and compactness, these confirmed relics present a rare opportunity to study such environments — allowing us to probe them in the X-ray domain as well.

In this context, \citet{werner2018digging} studied the extended, hot ISM in Mrk1216 and PGC032873 with \textit{Chandra} observations. Likewise, \citet{buote2019extremely} applied hydrostatic equilibrium models to their X-ray data to estimate the dark matter content, baryon fraction, and thermodynamic structure of the ISM. These works demonstrated the diagnostic power of X-ray morphology and thermodynamic properties.

Moreover, the hot ISM can be studied in the framework of chemical evolution and enrichment scenarios (e.g. early enrichment) \citep{werner2013uniform, mantz2017metallicity, biffi2017history}. Furthermore, the fact that relics have avoided the dry-merger phase offers a unique opportunity to isolate the effects of dynamical processes, such as mergers, on galaxy evolution.

In this study, we investigate the ISM of the isolated relic galaxy Mrk1216 to gain insights into its chemical enrichment history. We use archival \textit{XMM-Newton} observations, which offer improved spectral resolution compared to \textit{Chandra} data, allowing more precise abundance measurements. We compare our results with the radial abundance profiles of \citet{mernier2017radial}, based on a sample of 44 nearby cool-core galaxy clusters, groups, and ellipticals from the CHEERS (CHEmical Enrichment Rgs Sample) catalogue \citep{de2017cheers}.

This paper is organised as follows: Section \ref{mrk1216} provides background information on Mrk1216. Section \ref{obs_and_data_reduction} describes the data reduction, background treatment and spectral fitting methods. Section \ref{results} reports our findings, followed by Section \ref{discussion}, which explores their implications. Finally, Section \ref{conclusions} summarises the main results. Throughout this paper, we adopt the $\Lambda$CDM cosmology with a Hubble constant of $H_{0} = 70$ km s$^{-1}$ Mpc$^{-1}$, a matter density of $\Omega_{M} = 0.27$ and a cosmological constant of $\Omega_{\Lambda} = 0.73$. All quoted uncertainties are at the $1\sigma$ confidence level unless otherwise stated.

\section{Mrk1216}
\label{mrk1216}

Mrk1216 has been the subject of extensive multi-wavelength studies. Most notably, it exhibits structural peculiarities that distinguish it from other nearby massive Early-Type Galaxies (ETGs). In particular, Mrk1216’s halo density is a positive outlier in the $\Lambda$CDM $c_{200}/M_{200}$ relation, suggesting an unusually early formation time \citep{buote2019extremely}. Mrk1216 is likewise considered an isolated galaxy, with only two neighbouring galaxies identified within a 1-Mpc radius—both of which are more than two magnitudes fainter on the K-band \citep{yildirim2015mrk, ferre2017two, buote2019extremely}.

Mrk1216's total mass within $R_{500}$ is estimated to be $M_{500} = 4.5 \pm 0.5 \times 10^{12} M_{\odot}$, comprising the combined mass of its Black Hole (BH), stars, gas, and Dark Matter (DM), with a gas fraction of $f_{gas,500} = 0.058 \pm 0.008$ \citep{buote2019extremely}. The stellar, gas, and DM properties of Mrk1216 exhibit striking similarities to those of the nearby fossil group NGC6482 \citep{yildirim2017structural, buote2019extremely}. This resemblance, together with its extended X-ray emission, suggests that Mrk1216 may represent the central galaxy of a fossil group that subsequently formed around it \citep{buote2019extremely}.

The effective radius ($R_e$) of Mrk1216 is $2.3 \pm 0.1$ kpc, with a stellar velocity dispersion of $368 \pm 3$ km/s \citep{ferre2017two}.  Additionally, Mrk1216’s mean mass-weighted stellar age is $12.8 \pm 1.5$ Gyr, with a metallicity of $0.259 \pm 0.052$ dex \citep{ferre2017two}. Its stellar mass within $4R_e$ is $2.0 \pm 0.8 \times 10^{11} M_{\odot}$ \citep{ferre2017two}. Mrk1216’s morphology, dynamics and density profile closely resemble those of massive galaxies at $z>2$, distinguishing it from massive ETGs at $z \sim 0$ and corroborating its relic nature \citep{ferre2017two, werner2018digging}. As such, Mrk1216, in addition to being classified as both a Massive Relic Galaxy (MRG) and an Isolated Compact Elliptical Galaxy (IsoCEG), appears to be the dominant constituent of the possible fossil group surrounding it.

\citet{buote2019extremely} found no evidence of strong internal AGN outbursts in Mrk1216. In light of other findings (e.g., temperature, entropy, $t_c/t_{ff}$ profiles), they concluded that the galaxy exhibits gentle, precipitation-regulated AGN feedback. These characteristics make Mrk1216 an ideal candidate for inspecting the effects of passive AGN evolution.

\section{Observation and Data Reduction} 
\label{obs_and_data_reduction}
\label{sec:obs_red}

We analysed archival \textit{XMM-Newton} \citep{jansen2001} observations of Mrk1216 (RA:$08^{h}28^{m}47.11^{s}$, DEC:$-06^{d}56^{m}24.5^{s}$), conducted on 01 November 2018 (Observation ID 0822960201) with a total exposure of 109.7 ks.

\begin{figure}
    \centering
    \includegraphics[width=\linewidth]{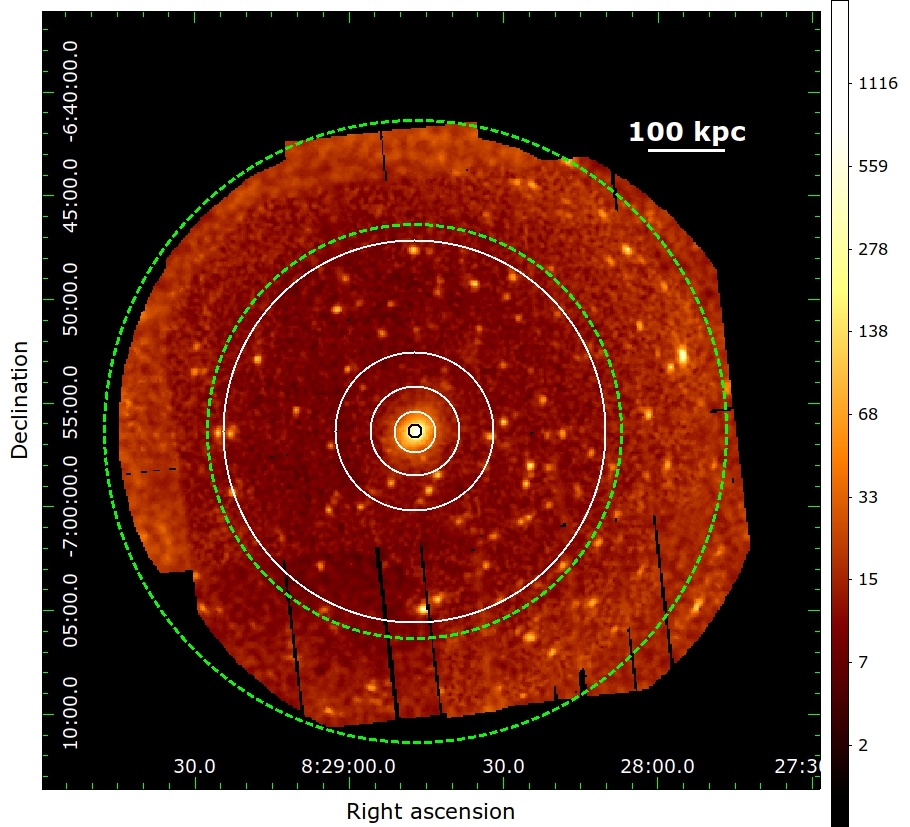}
    \caption{Adaptively smoothed and combined image of Mrk1216. The solid white annuli mark the regions selected for analysis, extending out to $R_{500}$. The central region is coloured black for visual clarity. The green dashed annulus indicates the background region.}
    \label{fig:Mrk1216}
\end{figure}

We used \textit{XMM-Newton} EPIC (MOS1, MOS2, and pn) and RGS instruments to analyse Mrk1216. While the EPIC observation was used to conduct imaging analyses and to extract spectra of selected annuli, as shown in Figure \ref{fig:Mrk1216}, the RGS was used to extract a high-resolution spectrum of the central area. The analysis procedure, methods employed and software used are described in detail in the following sections.

\subsection{EPIC Analysis} \label{epic_analysis}

We used the Extended Source Analysis Software (ESAS), integrated into the Science Analysis System (SAS) version 19, to process the \textit{XMM-Newton} data. Following the procedures outlined by \citet{snowden2014cookbook}, we ran \textit{emchain} on MOS event files and \textit{epchain} on pn event files for initial filtering and calibration. Faulty pixels were removed by filtering with the \texttt{FLAG == 0} condition. For MOS, we selected events with one to four pixels (PATTERN $\leq 12$), while for pn, only single- and double-pixel events (PATTERN $\leq 4$) were retained. We then performed \textit{mos-filter} and \textit{pn-filter} to identify intervals contaminated by soft protons (SPs), obtaining uncontaminated good time intervals (GTIs). 

Point sources were detected using the \textit{wavdetect} tool in the \textit{CIAO} software package \citep{fruscione2006ciao}. A \textit{mask} file incorporating the detected sources was created and applied in ESAS tasks to exclude them from the data. After cleaning, spectra were extracted using \textit{mos-spectra}, \textit{mos\_back}, \textit{pn-spectra}, and \textit{pn\_back}.

\subsection{Surface Brightness Profile} \label{sur_bri_profile}

Following the reduction steps, we extracted the surface brightness profile of Mrk1216. Such profiles describe how the ISM brightness changes radially, from the galaxy centre to the outskirts, allowing radial variations to be modelled. They also identify the radius where the ISM emission drops to the background level, providing the spatial extent of detectable emission. We fitted a $\beta$ model \citep{cavaliere1976x} to represent the ISM emission and a constant to represent background emission. Fit results are presented in Section \ref{results}.

\subsection{Spectral Analysis} \label{spectral_analysis}

The observed spectra include multiple components of different origins and characteristics. To properly model the galaxy emission, each component must be carefully treated. These components can be grouped as: Non-X-ray Background (NXB), Cosmic X-ray Background (CXB), and Source Emission (SE), described below.

\subsubsection{NXB (Non-X-ray Background)}

NXB arises when energetic particles interact with the detector or its surroundings, producing fluorescent X-rays that subsequently strike the detector \citep{kuntz2008epic}. NXB consists of three components: fluorescent instrumental lines in MOS and pn detectors, quiescent particle background (QPB), and residual soft-proton (SP) contamination. Fluorescent lines were modelled as Gaussian functions with given energies, as described in \citet{snowden2014cookbook}. 

QPB spectra were generated with ESAS extraction tasks and modelled with multiple power laws plus additional Gaussians as required. For each spectral region, we calculated the fit parameters and kept them fixed in the main model.

Although SP contamination is mitigated by light curve filtering during data reduction, as mentioned in Section \ref{epic_analysis}, we included a broken power law component to account for any potential residual SP contamination as described in \citet{snowden2014cookbook}. During the fitting process, we initially set the normalisations of the residual SP components to zero. Once a stable fit was achieved, we set them free to account for any residual contamination. 

Since these NXB components did not originate from X-ray photons, standard response files are not suitable for their modelling. Instead, we used diagonal response files provided in the ESAS Calibration Database (CALDB) as recommended by \citet{snowden2014cookbook}.

\subsubsection{CXB (Cosmic X-ray Background)}

CXB has three components—LHB (Local Hot Bubble), GH (Galactic Halo) and UPS (Unresolved Point Sources) \citep{de20042, kuntz2008epic}. In order to model these three components, we extracted spectrum for a 600–900 arcsec region as background area, shown in Figure \ref{fig:Mrk1216} as white dashed annuli. We also used the ROSAT All-Sky Survey (RASS) spectrum to fit our background spectra simultaneously with it. The RASS spectrum was obtained from the X-ray background tool provided by \textit{HEASARC}’s (High Energy Astrophysics Science Archive Research Center) website\footnotemark{}. We chose an annulus with radii of 1$^\circ$ and 2$^\circ$, centring the Mrk1216.

LHB is represented by a thermal APEC model with a fixed temperature value of 0.11 keV. GH is described by two APEC thermal emission models to account for hot and cold phases. UPS is modelled with a power law with a fixed index of ($\Gamma=1.45$). All thermal models’ metal abundance values are fixed at 1.0 and the redshift value is set to zero. Finally, additional absorption model is added to the GH and UPS components. 

After simultaneously fitting all background components to the RASS and EPIC spectra, the resulting CXB parameter values were used in the main model for the source regions. These values are listed in Table \ref{tab:bkg_table}.

\footnotetext[1]{heasarc.gsfc.nasa.gov/cgi-bin/Tools/xraybg/xraybg.pl}

{\renewcommand{\arraystretch}{1.2}
\begin{table}
	\centering
	\caption{The Cosmic X-ray background (CXB) fit parameters.}

	\begin{tabular} {@{}lcc@{}}
		\hline\hline
		Component & Parameter & Value\\
		\hline 
		LHB & kT (keV) & 0.11\\
		    & Norm (cm$^{-5}$) & 2.03 $\times$ 10$^{-7}$\\ \hline
	    GH$_{Cold}$ & kT (keV) & 0.10\\
	        & Norm (cm$^{-5}$) & 4.22 $\times$ 10$^{-6}$\\ 
	    GH$_{Hot}$ & kT (keV) & 0.30\\
	        & Norm (cm$^{-5}$) & 2.56 $\times$ 10$^{-7}$\\ \hline
        UPS & $\Gamma$ & 1.45 \\ 
		    & Norm (cm$^{-5}$) & 8.78 $\times$ 10$^{-7}$\\ \hline
		\hline
\end{tabular}

\label{tab:bkg_table}
\end{table}}

\subsubsection{Source Emission}

Source emission comprises ISM and LMXB (low-mass X-ray binary) contributions. LMXBs create a hard continuum, typically as point sources, with nearly universal cumulative spectra \citep{irwin2003x, ji2009elemental}. While detected point sources were removed, unresolved LMXB emission can remain \citep{kim2011metal}. Although minor for Mrk1216 \citep{buote2019extremely}, it was included for completeness. The source emission was therefore modelled with a thermal APEC for ISM and a power law with fixed $\Gamma = 1.6$ for LMXBs \citep{irwin2003x}, both absorbed by Galactic $n_H$.

Accurate ISM modelling is crucial for reliable abundance measurements. A single-temperature plasma assumption can bias results, particularly producing the Fe-bias effect \citep{buote1998x}. To mitigate this, we adopted the VGADEM multi-temperature model in XSPEC \citep{xspec1999}, which assumes a Gaussian distribution of emission measures. The model yields mean ($\mu_{kT}$) and width ($\sigma_{kT}$) of the temperature distribution, as well as abundances of multiple elements relative to Solar (He, C, N, O, Ne, Na, Mg, Al, Si, S, Ar, Ca, Fe, Ni).

\subsection{RGS Analysis} \label{rgs_analysis}

High-resolution spectra were extracted from RGS following the official guidelines\footnote{https://heasarc.gsfc.nasa.gov/docs/xmm/abc/abc.html}. The extraction region was set to 40 arcsec width, matching the 20-arcsec radius central region used in EPIC analysis.

\subsection{Spectral Fitting} \label{spectral_fitting}

\begin{figure*}
    \centering
    \includegraphics[width=1.0\linewidth]{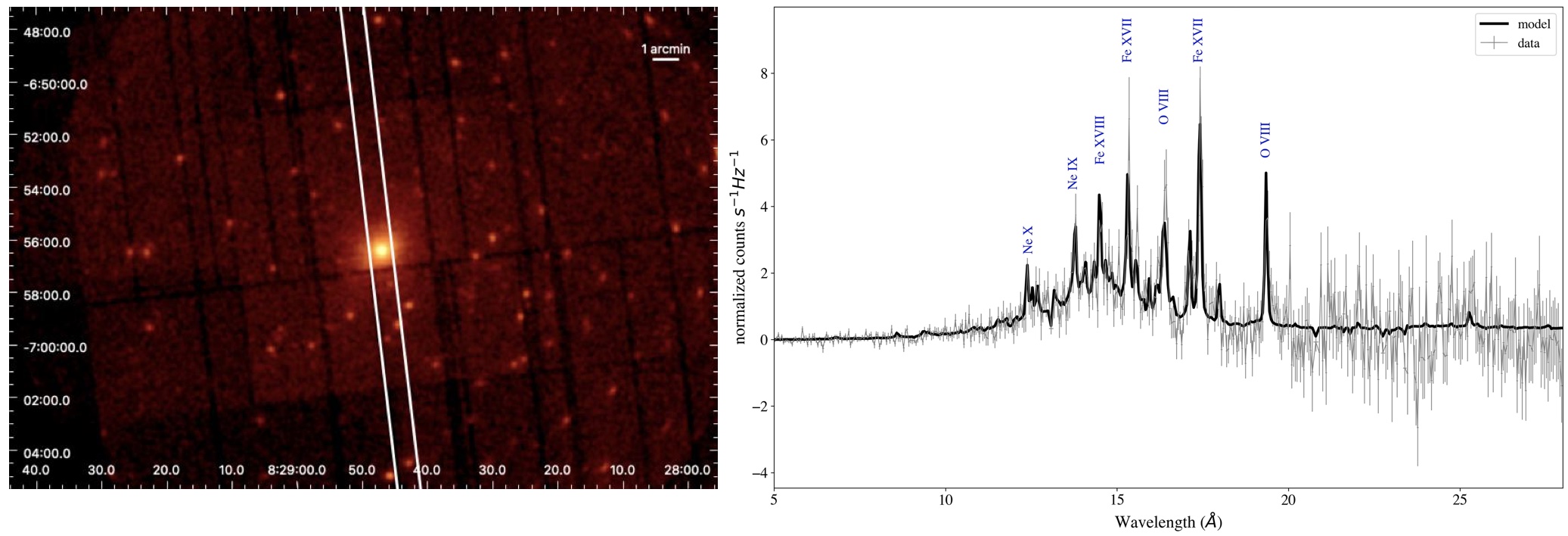}
    \caption{\textit{Left}: Adaptively smoothed and combined image of Mrk1216. White lines show the RGS spectral region. \textit{Right}: Spectral fit of the RGS region.}
    \label{fig:Mrk1216_RGS}
\end{figure*}

\subsubsection{EPIC}\label{spectral_fitting_epic}

Spectral fitting was performed using XSPEC 12.11.1 \citep{xspec1999} and AtomDB 3.0.9 \citep{apec_2001,foster2012updated}. MOS spectra were restricted to 0.3–6.5 keV, and pn to 0.4–6.5 keV. We used C-statistics \citep{cash1979}, meaning no binning or background subtraction was applied, as required. Elemental abundances were calculated relative to Solar values using the \textit{LPGS} table \citep{lodd2009}.

The EPIC spectral model is given in Eq. \ref{eq:epic_spectral_model}. The \textit{FL} term represents MOS and pn fluorescent lines, modelled with Gaussian components \citep{snowden2014cookbook}. The first \textit{Const} accounts for cross-calibration offsets among MOS1, MOS2, and pn, and the second for differing extraction areas, scaled to match RASS following \citet{snowden2014cookbook}. Galactic absorption was modelled with PHABS using $n_H$ from \citet{ben2016hi4pi}. Since NXB is particle-based, it was modelled separately with ESAS diagonal responses \citep{snowden2014cookbook}. 

\begin{equation}
    S = [FL + Const \times Const \times Abs(CXB + Source)] + [NXB]
    \label{eq:epic_spectral_model}
\end{equation}

\subsubsection{RGS}

The same tools and statistics were used for RGS fitting, restricting spectra to 7–22 Å. O, Ne, and Fe abundances were left free in the VGADEM component. Compared to EPIC, RGS fitting was straightforward, requiring only the plasma emission plus Galactic absorption (TBABS). The RGS spectral region and model are shown in Figure \ref{fig:Mrk1216_RGS}.

\begin{figure}
    \centering
    \includegraphics[width=\linewidth]{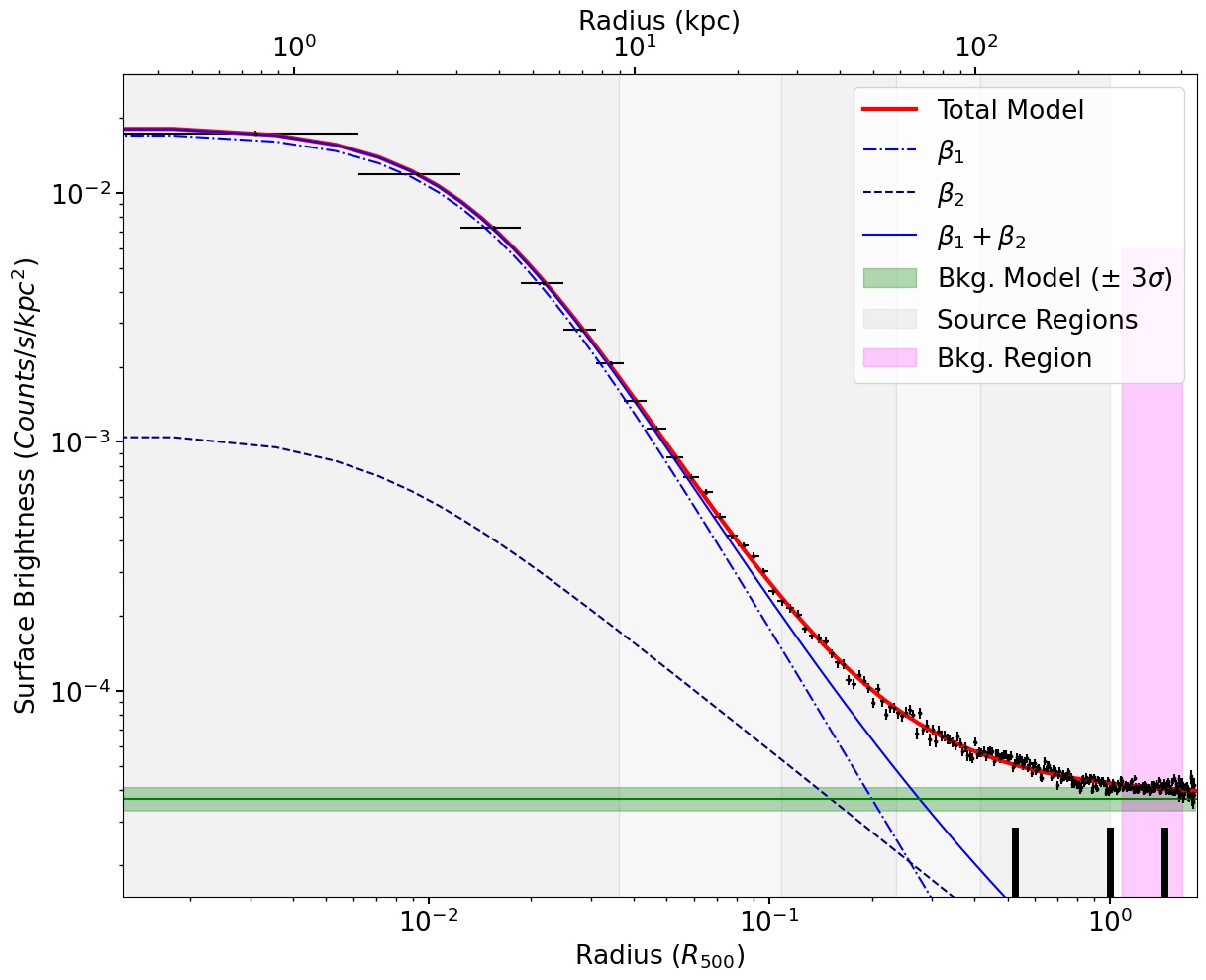}
    \caption{Black data points show the surface brightness profile of Mrk1216. The red line shows the best-fit model, consisting of $\beta_1$, $\beta_2$, and constant background terms. The constant background is plotted as a green line with light-green $3\sigma$ uncertainty. Six annuli were selected for spectra, with the sixth for background. Alternating grey areas mark annuli, the pink region marks the background annulus. Black lines at bottom right mark $R_{2500}$, $R_{500}$, and $R_{200}$ from left to right.}
    \label{fig:Mrk1216_sur_bri}
\end{figure}

\subsection{SNe Ratio}

We used the \texttt{SNeRatio} code \citep{erdim2021relative} to estimate the relative contributions of different supernova types to ISM enrichment. The code models observed ISM abundances with yield sets from SNIa and SNcc progenitors (summarised in Table \ref{table:yieldmods}). The output is the SNIa fraction relative to the total supernova population (SNIa + SNcc), obtained by comparing X-ray derived ISM abundances with the theoretical yields.

{\renewcommand{\arraystretch}{1.2}
\begin{table}
\caption{Adopted Yields in {\texttt{SNeRatio}} tool.}
\centering 
\begin{threeparttable}[b]
\begin{tabular}{@{}lccc@{}} 
\hline\hline 
Reference & Reference & Initial & SN \\
 & Table\tnote{a} & Metallicity\tnote{b} & Type \\
\hline 
\citet{seitenzahl2013three} & Table 2 & -- & SNIa \\\hline
\citet{Fink2013} & Table B1 & -- & SNIa \\\hline
\citet{nomoto2013nucleosynthesis} & Table\tnote{c} & 0.0   & SNcc \\
                                                     &                               & 0.001 &  \\
                                                     &                               & 0.004 &  \\
                                                     &                               & 0.008 &  \\
                                                     &                               & 0.02  &  \\
                                                     &                               & 0.05  &  \\
\hline
\end{tabular}

\begin{tablenotes}
     \item[a] This column refers to cited table numbers.
     \item[b] This column is given in units of Z$_{\odot}$.
     \item[c] {Available at: \url{http://star.herts.ac.uk/chiaki/works/YIELDCK13.DAT}}
\end{tablenotes}
\end{threeparttable}
\label{table:yieldmods}
\end{table} }

\begin{equation}
R_{(Ia)} = \frac{SNIa}{SNIa + SNcc}.   
\label{eq:RIa}
\end{equation}

\section{Results} \label{results}

In this work, we used X-ray spectral analyses to investigate the chemical properties of the isolated elliptical relic galaxy Mrk1216. The spectral analysis covered five concentric circular annuli and a single circular region encompassing the area within $R_{500}$, together with an additional background region beyond $R_{500}$. After calculating the background emission parameters, the ISM emission was modelled for each region as outlined in Section \ref{obs_and_data_reduction}.

During spectral modelling, an unidentified excess emission was detected at $\sim$1.2 keV in the central two regions. Curiously, \citet{buote2019extremely} reported the same excess emission in \textit{Chandra} observations of Mrk1216 and proposed possible explanations for this phenomenon. That two different observatories detected the same excess suggests that this is not instrument-related incident. Nevertheless, identifying the origin of this excess emission requires further investigation, which is beyond our scope. To mitigate potential biases introduced by this excess, the 1.17–1.25 keV energy range was removed from the spectra for the regions in question. This exclusion was applied only after careful consideration, and we verified that it did not affect the modelling of the continuum or elemental abundances. 

We also noted the presence of slightly S-shaped residuals in the modelled spectra of the $0-R_{500}$ region (0-248 kpc). Each annulus yielded statistically reliable results independently, suggesting that the issue most likely arises from a substantial mixture of multi-phase plasma—i.e., plasma encompassing a wide range of temperatures and metal abundances—probably due to the relatively large spatial coverage. A similar trend was reported by \citet{kim2011metal}. Therefore, spectral analysis results for $0-R_{500}$—originally intended to represent the overall averages of the galaxy—are likely affected and should be interpreted with caution. That said, we did not observe this trend in any of the other regions.

The analysis of the selected regions includes radial profiles of X-ray luminosity, plasma temperatures, elemental abundances and SNIa ratios. Several of these properties have previously been examined using \textit{Chandra} X-ray observations of Mrk1216 \citep{buote2018luminous, buote2019extremely, werner2018digging}. Among these, we primarily compared our results with \citet{buote2019extremely}, as their study used the most recent and deepest \textit{Chandra} observations.

While the higher spatial resolution of the \textit{Chandra} satellite allows the central areas to be examined in greater detail, the higher spectral resolution of the \textit{XMM-Newton} satellite afforded us the ability to measure element abundance values with more precision. Moreover, \textit{XMM}'s wider field of view made it possible to extend the analysis up to $R_{500}$ for a subset of the parameters. However, due to decreasing data quality at larger radii, not all parameters could be constrained out to $R_{500}$ — some could be measured up to this radius, while others were only accessible within smaller radii.

\subsection{Surface Brightness Profile}

To study the ISM emission, we followed the procedure described in Section \ref{epic_analysis}, using a mask file to remove point sources. We derived a $\beta$ model \citep{cavaliere1976x} to model the ISM emission and a constant model to account for any background emissions. However, the single-$\beta$ model was failed to adequately reproduce the data. The fit was greatly improved by adding an additional $\beta$ model, yielding a double-$\beta$ model plus a constant background. Whereas the first $\beta$ model corresponds to the main emission component of the galaxy and has a brighter peak with a slightly steeper decline, the second $\beta$ model is fainter in the centre but begins to dominate from approximately the fourth radial bin onward ($\sim 65 kpc$). The $r_{c1}$ and $\beta_1$ values ($3.41\pm0.34$ kpc and $0.55\pm0.02$) are consistent with \citet{buote2019extremely}’s  analysis results of  $6.2_{-1.3}^{+2.0}$ arcsecs ($2.8_{-0.6}^{+0.9}$ kpc) and $0.52_{-0.02}^{+0.03}$, respectively. Fit parameters are listed in Table \ref{tab:table_sur_bri} and the modelled surface brightness profile is shown in Figure \ref{fig:Mrk1216_sur_bri}. 

\begin{table}
    \caption{Fit parameters of surface brightness profile. (Two $\beta$ models and a constant background model.)}
    \begin{tabular}{c|r|l|l} 
      \hline
      \textbf{Parameter} & \textbf{Value} & & \textbf{Unit}\\
      \hline
      \textbf{$S_1$} & $2.77 \pm 0.43$ & $(x10^{-3})$ &  ($cts/s/kpc^2$)\\
      \textbf{$r_{c1}$} & $3.41 \pm 0.34$ & & ($kpc$)\\
      \textbf{$\beta_1$} & $0.55 \pm 0.02$ & & - \\
      \hline
      \textbf{$S_2$} & $0.17 \pm 0.50$ & $(x10^{-3})$ & ($cts/s/kpc^2$)\\
      \textbf{$r_{c2}$} & $1.72 \pm 3.06$ & & ($kpc$)\\
      \textbf{$\beta_2$} & $0.35 \pm 0.06$ & & -\\
      \hline
      \textbf{$Bkg.$} & $5.91 \pm 0.21$ & $(x10^{-6})$ & ($cts/s/kpc^2$)\\
      \hline
    \end{tabular}
    \label{tab:table_sur_bri}
\end{table}

To define the spectral regions, we set a hard radial limit at the point where the brightness drops to the modelled background level within $3\sigma$ uncertainty. Since we noticed that the $R_{500}$ value of Mrk1216 lies with this boundary, we set the radial limit at $R_{500}$ ($\sim 248 kpc$) to facilitate comparisons with other studies.

We divided the $R_{500}$ region into five concentric annuli for spectral analysis. We also defined a background annulus beyond the $R_{500}$, with inner and outer radii of $268.8,\mathrm{kpc}$ and $403.2,\mathrm{kpc}$, respectively, in order to model the background emission. This region was placed as far from the centre as possible to produce a more accurate spectral representation of the background emission. Source regions are shown as solid white circles and the background region as green-dashed annuli (Figure \ref{fig:Mrk1216}). For reference, the short black vertical lines at the bottom right of Figure \ref{fig:Mrk1216_sur_bri} represent Mrk1216’s radial values corresponding to $R_{2500}$ ($\sim 130 kpc$), $R_{500}$ ($\sim 248 kpc$) and $R_{200}$ ($\sim 358 kpc$), as reported by \citep{buote2019extremely}.

{\renewcommand{\arraystretch}{1.6}
\begin{table*}
	\centering
	\caption{Spectral fit results of Mrk1216.}

    \resizebox{\textwidth}{!}{%
	\begin{tabular} {@{}ccccccccccc@{}}
		\hline\hline
		Instrument & Radius (kpc) & $kT_{\mu} (keV) $ & $kT_{\sigma}$ (keV) & O & Ne & Mg & Si & S & Fe & Stat./dof \\
		\hline\hline

        RGS & 0-8.96 & $0.800_{-0.019}^{+0.021}$ & $0.232_{-0.051}^{+0.054}$ & $0.45_{-0.11}^{+0.16}$ & $0.35_{-0.14}^{+0.19}$ & - & - & - & $0517_{-0.105}^{+0.155}$ & 1729.21/1468 \\ \hline

		EPIC & 0-8.96 & $0.770_{-0.004}^{+0.004}$ & $0.155_{-0.022}^{+0.019}$ & - & - & $1.11_{-0.11}^{+0.14}$ & $0.80_{-0.08}^{+0.11}$ & $1.06_{-0.20}^{+0.24}$ & $0.891_{-0.061}^{+0.082}$ & 2638.47/3639 \\ \cline{2-11}
  
		 & 8.96-26.88 & $0.686_{-0.007}^{+0.007}$ & $0.150_{-0.029}^{+0.025}$ & - & - & $0.77_{-0.09}^{+0.08}$ & $0.56_{-0.08}^{+0.09}$ & $0.71_{-0.23}^{+0.28}$ & $0.630_{-0.050}^{+0.059}$ & 3657.81/3640 \\ \cline{2-11}
  
         & 26.88-58.24 & $0.655_{-0.013}^{+0.009}$ & $0.122_{-0.046}^{+0.047}$ & - & - & $0.57_{-0.12}^{+0.17}$ & $0.47_{-0.13}^{+0.15}$ & - & $0.592_{-0.073}^{+0.117}$ & 4032.11/3690 \\ \cline{2-11}
        
         & 58.24-103.04 & $0.641_{-0.016}^{+0.020}$ & $0.054_{-0.054}^{+0.077}$ & - & - & $0.56_{-0.23}^{+0.19}$ & $0.35_{-0.33}^{+0.34}$ & - & $0.543_{-0.052}^{+0.048}$ & 3975.01/3690 \\ \cline{2-11}

         & 103.04-248.19 & $0.608_{-0.069}^{+0.062}$ & $0.010_{-0.010}^{+0.208}$ & - & - & - & - & - & $0.129_{-0.042}^{+0.048}$ & 3918.67/3690 \\ 
         \cline{2-11}\cline{2-11}

         & 0-248.19 & $0.713_{-0.003}^{+0.006}$ & $0.123_{-0.012}^{+0.012}$ & - & - & $0.52_{-0.04}^{+0.04}$ & $0.22_{-0.05}^{+0.05}$ & $0.41_{-0.22}^{+0.22}$ & $0.375_{-0.004}^{+0.003}$ &  3879.53/3689 \\ 
         
         \hline\hline

    \end{tabular}%
    }

\label{tab:fit_table}
\end{table*}}

\subsection{Temperature Profile}

We observed a slight radial decline in $\mu_{kT}$ from $\sim$0.77 keV to $\sim$0.61 keV out to $R_{500}$. This decrease is significant in the first three regions, while the last two are within uncertainties ($\sigma_{kT}$). The central region has the broadest temperature distribution ($\sigma_{kT} \sim 0.16$). Although uncertainties increase with radius, the $\sigma_{kT}$ values generally decrease, reaching near zero in the outermost region. This suggests that a single-temperature approximation may be valid there, although large uncertainties make this conclusion uncertain. Elevated background and low source counts likely contributed to these uncertainties.

\citet{buote2019extremely} used \textit{Chandra} to measure temperatures out to $\sim$110 kpc, finding a central peak followed by a decline between $\sim$0.6–1.0 keV (Table 6 of the reference paper). This corresponds to our innermost four bins ($\sim$103 kpc). Although the two datasets used different instruments and models (single- vs. multi-temperature), the average temperatures are broadly consistent.

\begin{figure}
    \centering
    \includegraphics[width=\linewidth]{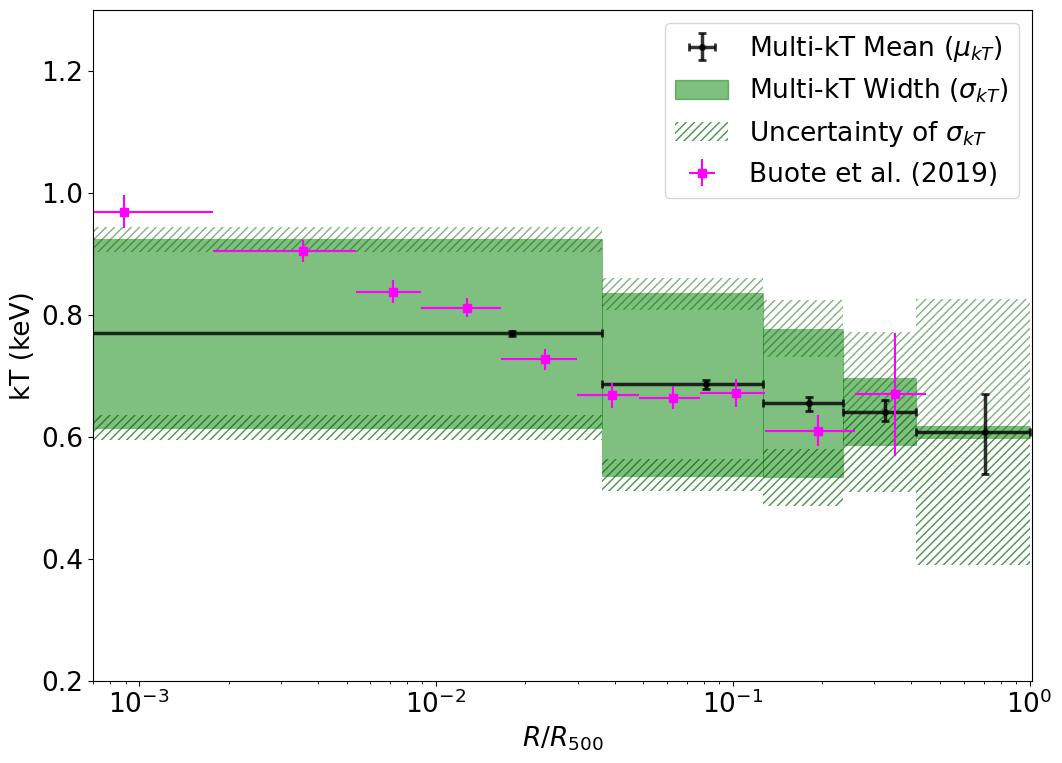}
    \caption{Radial temperature profile of Mrk1216. The derived multi-temperature plasma emission model (VGADEM) calculates the mean ($\mu_{kT}$) and standard ($\sigma_{kT}$) deviation of temperature values for a specific region. Temperature values from \citet{buote2019extremely} are shown in magenta for comparison.}
    \label{fig:Mrk1216_kT}
\end{figure}

\subsection{Abundance Profile}

\begin{figure*}
    \centering
    \includegraphics[width=\linewidth]{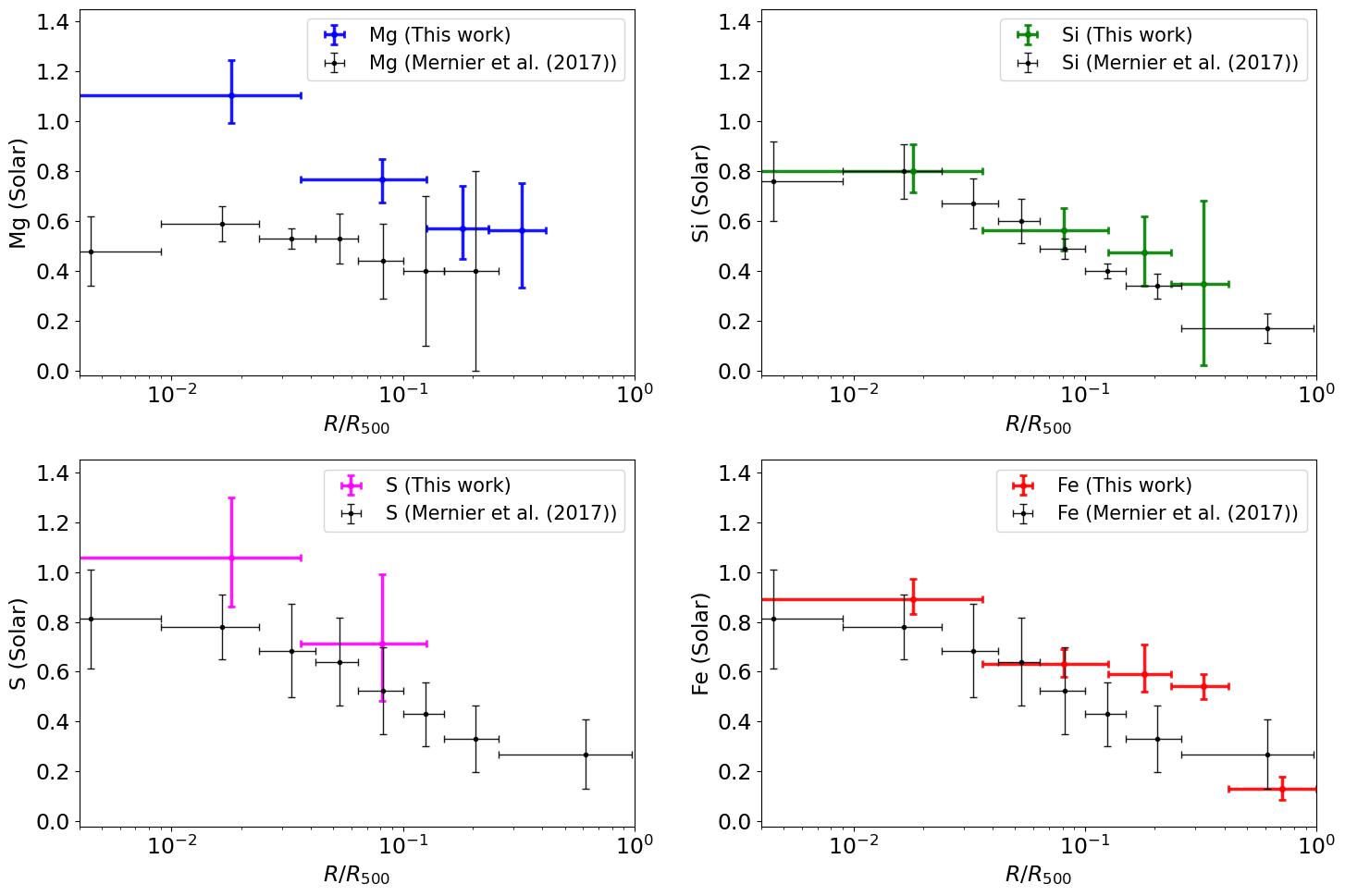}
    \caption{Radial abundance profiles (Mg, Si, S and Fe) of Mrk1216 and groups average from \citet{mernier2017radial}.}
    \label{fig:Mrk1216_mg_si_fe_mernier}
\end{figure*}

We successfully measured Fe abundances out to $R_{500}$. Mg and Si were measured up to the fourth annulus ($\sim0.42R_{500}$), and S in the two central regions ($\sim0.11R_{500}$). The resulting abundance profiles are presented in Figure \ref{fig:Mrk1216_mg_si_fe_mernier}. RGS allowed additional measurements of O, Ne, and Fe in the central $\sim$9 kpc. Table \ref{tab:fit_table} lists all abundances.

Si-K fluorescence contamination was detected in annuli three and four. As noted by \citep{snowden2014cookbook}, this line emission is present only in the MOS1 and MOS2, but not in the pn detector. To avoid biases, Si abundances in these regions were taken exclusively from pn data. 

With respect to the EPIC measurements, the Mg, Fe, and Si profiles all appear to peak centrally and then flatten out towards the outer regions. However, given the large error bars, it is difficult to reliably quantify this trend. The Fe profile also drops sharply in the final region, beyond the flattening.

\begin{figure*}
    \centering
    \includegraphics[width=\linewidth]{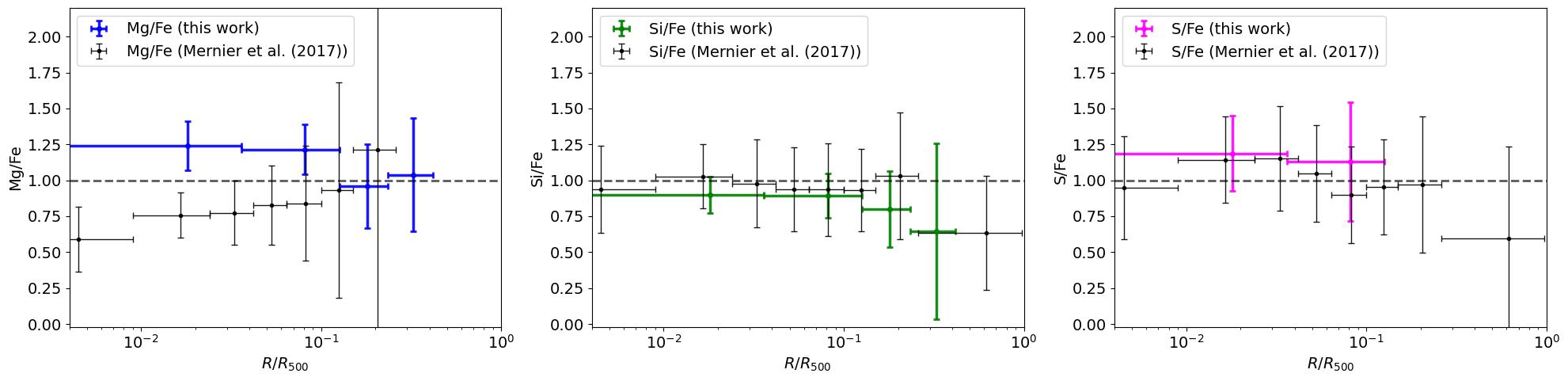}
    \caption{Radial abundance ratio profiles ([Mg/Fe], [Si/Fe] and [S/Fe]) of Mrk1216 and groups average from \citet{mernier2017radial}.}
    \label{fig:Mrk1216_mgRatio_siRatio_mernier}
\end{figure*}

Because abundance ratios are more robust tracers of enrichment \citep{kim2011metal}, we calculated Mg/Fe, Si/Fe, and S/Fe. Mg/Fe starts above Solar, decreasing toward unity. Si/Fe is consistent with Solar within uncertainties, while S/Fe is poorly constrained but also consistent with Solar. Ratios are shown in Figure \ref{fig:Mrk1216_mgRatio_siRatio_mernier}.

With a large sample size of 44 nearby objects, \citet{mernier2017radial} is one of the most comprehensive and up-to-date studies on the chemical enrichment of elliptical galaxies, groups and clusters. Said study divided the sample into two categories based on temperature, given its correlation with mass. The first category consists of galaxy clusters above 1.7 keV and the second of elliptical galaxies and galaxy groups below 1.7 keV. The analysis results of Mrk1216 can be compared with the second category (<1.7 keV) to ascertain whether it follows the expected average behaviour of the aforementioned sample or exhibits distinct features linked to its unique evolutionary history. We should note that both studies adopt the same solar table (LPGS, \citet{lodd2009}), which eliminates potential systematic differences between solar tables when comparing abundance values.

We compared the abundance values of Mg, Si, S, and Fe with those of \citet{mernier2017radial} (Figure \ref{fig:Mrk1216_mg_si_fe_mernier}). Our Mg profile shows a central peak, unlike their nearly flat distribution, with higher values in the innermost two regions, while the outer bins are consistent. The Si profiles are in good agreement. Our S profile, limited to the two central bins, is also consistent. For Fe, both studies reveal a similar overall trend, though our values are higher in the third and fourth regions and drop sharply in the outermost bin.

In terms of abundance ratios (Figure \ref{fig:Mrk1216_mgRatio_siRatio_mernier}), the main difference lies in Mg/Fe: our profile starts with a super-solar value (significant at $\sim1\sigma$) and decreases with radius, whereas theirs begins at a lower ratio ($\sim0.5$) and increases outward. The Si/Fe and S/Fe ratios, however, are in good agreement between the two studies. The effect of the Fe drop in the outermost bin could not be assessed since other elements were not measurable at that radius.

\subsection{SNIa Ratio Profile}

\begin{figure}
    \centering
    \includegraphics[width=0.70\linewidth]{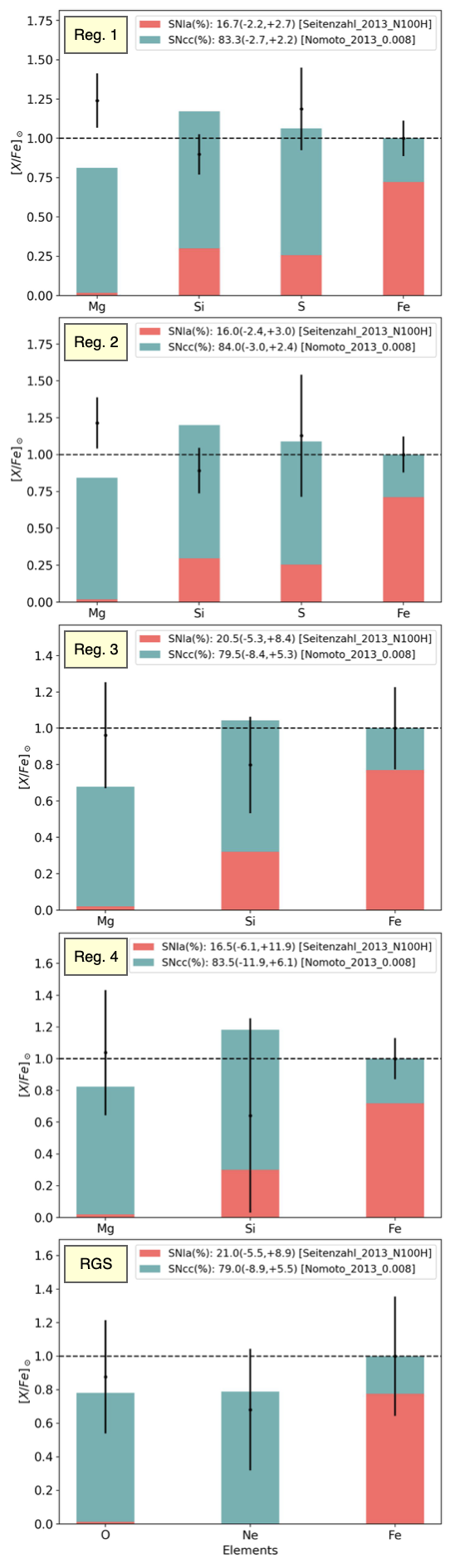}
    \caption{SNeRatio fit results for the four central EPIC regions and for the RGS region. The black data points represent the measured [X/Fe] ratios with their associated uncertainties. The vertical bars show the model prediction for the [X/Fe] ratios, with red segments corresponding to the SNIa contribution and green segments to the SNcc contribution. The black horizontal dashed lines indicate the solar ratios.}
    \label{fig:Mrk1216_sneratio_fits}
\end{figure}

SNIa ratio profiles are valuable tools for examining the chemical evolution of galaxies, groups, and clusters, as both the ratio itself and its radial variation provide important insights into their evolutionary history. We used the \texttt{SNeRatio} code \citep{erdim2021relative} to estimate the relative SNIa contribution up to $\sim0.42 \, R_{500}$. Fits included Mg/Fe, Si/Fe, and S/Fe in the central regions, and Mg/Fe, Si/Fe in the outer regions. O/Fe and Ne/Fe from RGS were also fitted. Results are shown in Figure \ref{fig:Mrk1216_sneratio_fits}.

\begin{figure}
    \centering
    \includegraphics[width=\linewidth]{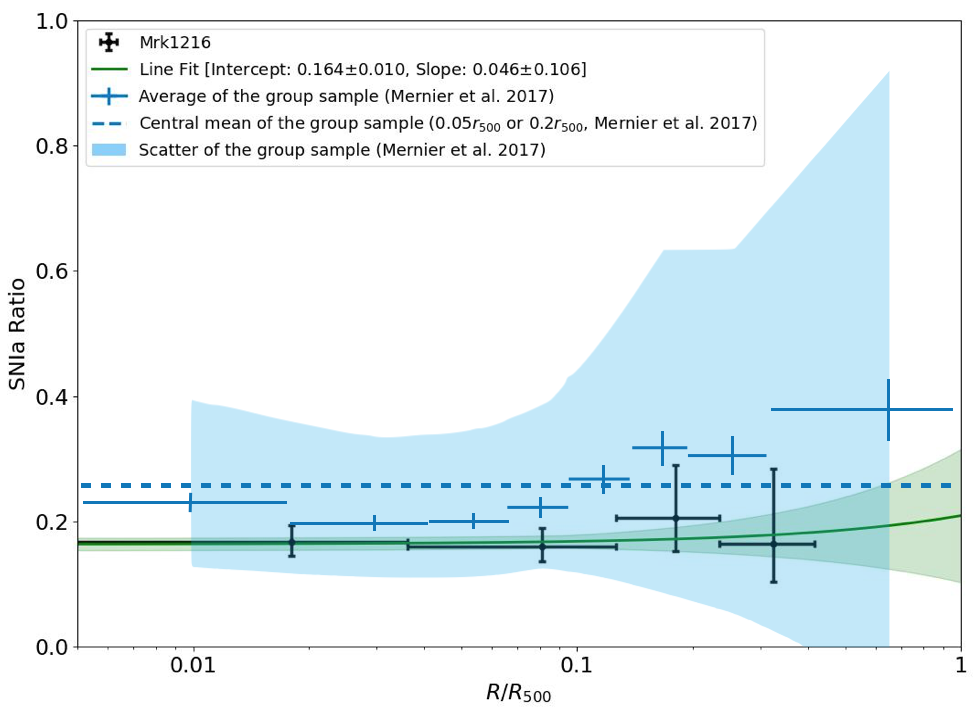}
    \caption{Radial SNIa ratio profile of Mrk1216 (black data points) compared with the groups’ average from \citet{mernier2017radial} (blue data points). The solid green line shows the line fit whereas the shaded green area shows the uncertainty of this fit. The intercept and slope parameters of this line fit are written in the label. The dashed blue line shows the average central value ($0.05r_{500}$ or $0.2r_{500}$) and the shaded blue area shows the scatter of the uncertainties of the group sample from \citet{mernier2017radial}.}
    \label{fig:Mrk1216_snIa_const_fit}
\end{figure}

Systematic differences can arise with different yield models \citep{erdim2021relative}. To ensure consistency, we adopted the same parameters as \citet{mernier2017radial}: the $N100H$ SNIa yields \citep{seitenzahl2013three}, $Z0.008$ SNcc yields \citep{nomoto2013nucleosynthesis}, a Salpeter IMF (10–40 $M_\odot$) \citep{salpeter1955luminosity}, and the LPGS solar table \citep{lodd2009}. Fit results are in Table \ref{tab:SNeRatio_fit_results} and Figure \ref{fig:Mrk1216_snIa_const_fit}.

As illustrated in Figure \ref{fig:Mrk1216_sneratio_fits}, the yield model fit for regions one and two underestimates the Mg/Fe ratio and overestimates the Si/Fe ratio. To assess the impact of these discrepancies on the estimated SNIa ratio, we repeated the fits for both regions excluding the Mg/Fe. This resulted in  slightly higher SNIa ratios of $0.233_{-0.041}^{+0.056}$ and $0.250_{-0.055}^{+0.081}$ along with improved fit statistics ($\chi^2/dof$ = 1.73/1 and 0.56/1, respectively). Excluding Si/Fe from the fits further improved the statistics and significantly lowered the SNIa ratios to $0.103_{-0.018}^{+0.023}$ and $0.102_{-0.019}^{+0.025}$, with corresponding $\chi^2/dof$ values of $0.72/1$ and $0.45/1$. Possible explanations for these under- and overestimations are discussed in Section \ref{discussion}. Region three and four, by contrast, exhibit better fit statistics and align more closely to model expectations.

To measure radial behaviour, we fitted a linear model to the SNIa ratio profile. The resulting fit was nearly flat with a small slope of $0.046\pm0.106$ and an intercept of $0.164\pm0.10$, as shown in Figure \ref{fig:Mrk1216_snIa_const_fit}. For comparison, we repeated the fit with a constant model—a line with zero slope—to quantify the average SNIa ratio value for $\sim0.42 \, R_{500}$ radius, which we calculated to be $0.167\pm0.006$.

Figure \ref{fig:Mrk1216_snIa_const_fit} depicts the SNIa ratio profile of Mrk1216 from this work alongside the average profile of groups and ellipticals from \citet{mernier2017radial}. Our results are represented by black data points, whereas those of \citet{mernier2017radial} are shown in blue. The horizontal blue dotted line indicates the average central SNIa ratio for groups and ellipticals and the shaded blue area represents associated scatters (see the paper for details).

{\renewcommand{\arraystretch}{1.6}
\begin{table}
    \caption{SNeRatio fit results.}
    \centering
    \begin{tabular}{ccccc}
    \hline\hline
        Instrument & Radius (kpc) & SNIa Ratio & Stat./dof \\
        \hline\hline
        RGS & 0-8.96 & $0.210_{-0.055}^{+0.089}$ & 0.17/1 \\
        \hline
         & 0-8.96 & $0.167_{-0.022}^{+0.027}$ & 11.1/2 \\ \cline{2-4}
        EPIC & 8.96-26.88 & $0.160_{-0.024}^{+0.030}$ & 8.65/2 \\ \cline{2-4}
         & 26.88-58.24 & $0.205_{-0.053}^{+0.084}$ & 1.78/1 \\ \cline{2-4}
         & 58.24-103.04 & $0.165_{-0.061}^{+0.119}$ & 1.07/1 \\ 
        \hline\hline
    \end{tabular}
    \label{tab:SNeRatio_fit_results}
\end{table}
}

\section{Discussion} \label{discussion}

\subsection{Temperature Structure}

Previous studies using \textit{Chandra} observations have reported the temperature profile of Mrk1216 \citep{werner2018digging, buote2019extremely}. Thanks to its higher spatial resolution, \textit{Chandra} provides a more detailed view of the central temperature structure by employing narrower annuli. The results are consistent, revealing a centrally peaked radial temperature profile characterised by a negative gradient (see Figure \ref{fig:Mrk1216_kT}).

\citet{kim2020temperature} analysed 60 ETGs and proposed a classification of temperature profiles into six categories: hybrid-bump, hybrid-dip, double-break, positive, negative, or irregular (see their paper for detailed definitions). Based on its clear negative gradient, Mrk1216 falls into the negative class. Typically, negative-type ETGs are dynamically disturbed and resemble non-cool-core (NCC) clusters. However, Mrk1216 is a highly relaxed system with no signs of dynamical disturbance \citep{yildirim2015mrk, werner2018digging, buote2019extremely}, making it an outlier.

Interestingly, the sample in that study includes another exception: NGC6482. Despite being a relaxed fossil system \citep{khosroshahi2004old}, it also shows a negative temperature gradient. The similarity between these two fossil systems (Mrk1216 and NGC6482) compared with normal ETGs may provide valuable insight into heating and cooling mechanisms, highlighting the role of galaxies’ evolutionary histories. \citet{werner2018digging} proposed radio-mechanical AGN feedback as the main heating source for Mrk1216, identifying it as an outlier in both the black hole–bulge mass relation and chaotic cold accretion (CCA) \citep{gaspari2013chaotic}. Further discussion of heating mechanisms is beyond the scope of this paper.

\subsection{Metallicity Structure}

As noted above \citep[and references therein]{goulding2016massive}, the two-phase galaxy formation scenario posits that structures are formed via gas-rich (wet) and dissipative mergers ($z \gtrsim 2$) followed by a phase of passive evolution through gas-poor (dry), non-dissipative and collisionless mergers ($2 \gtrsim z \gtrsim 0$). The wet mergers characteristic of the first phase induce rapid and efficient episodes of star formation \citep{naab2009minor, oser2010two, werner2018digging, buote2019extremely}. During this process, $\alpha$-elements—such as Mg—synthesised in the cores of massive stars are expelled through core-collapse supernovae (SNcc). The short lifespan of these massive stars results in the rapid release and recycling of Mg, making it a reliable tracer for overall enrichment.

In contrast, type Ia supernovae (SNIa), which originate from the remnants of long-lived, low-mass stars,  produce mainly Fe-peak elements \citep{beverage2021elemental}. Given the time delay between SNcc and SNIa, rapid star formation leads to an overabundance of $\alpha$-elements (from SNcc) relative to Fe-peak elements (from SNIa). As a result, enhanced $[\alpha/Fe]$ elemental abundance ratios serve as indicators of the intensity and timescale of star formation episodes \citep{thomas2005epochs, thomas2010environment, spolaor2010early}.

Red nuggets formed through these wet mergers, and their relics, are therefore expected to show high $[\alpha/Fe]$. Several optical spectroscopic studies have demonstrated this \citep{ferre2017two, spiniello2021inspire_a, spiniello2024inspire}. The abundance ratio derived from the X-ray-emitting plasma of Mrk1216 exhibits the same behaviour, showing a notably high [Mg/Fe] ratio.

Numerous studies have examined the relationship between mass and metallicity, commonly referred to as the mass–metallicity relation (MZR), finding that metallicity tends to increase with galactic mass \citep{lequeux1979reprint, tremonti2004origin}. This trend is often attributed to the greater efficiency with which massive systems enrich accreted gas \citep{spolaor2010early}. However, subsequent research has suggested that compactness may play a more significant role in this regard. Specifically, the depth of a galaxy’s gravitational potential governs its metallicity, such that, at fixed mass, more compact galaxies have steeper potential wells and consequently exhibit higher metallicities \citep{hoopes2007diverse, mcdermid2015atlas, ellison2018enhanced, maiolino2019re, barone2018sami, beverage2021elemental}.

The galaxies and groups in the CHEERS sample \citep{mernier2017radial}, against which our findings were compared, exhibit temperatures ranging from 0.5 to 1.7 keV, with a mean of approximately 0.95 keV and a scatter of 0.32 keV \citep{de2017cheers}. These temperatures served as a proxy for total mass. At $kT_{\mu,500}\approx0.71$, Mrk1216 lies below this average (Table \ref{tab:fit_table}). Given its compact nature as a relic, measuring elevated metal abundances in the X-ray band would lend support to the idea that compactness, rather than mass alone, drives higher metallicity. Only central Mg differs significantly, while other abundances agree within errors, yet this pattern appears broadly consistent with the proposed scenario.

Furthermore, the high central [Mg/Fe] value is in line with expectations for relic galaxies, where elevated [Mg/Fe] ratios reflect rapid early star formation and enrichment. These results support the classification of Mrk1216 as a relic system. To strengthen the generality of these findings, studying a larger sample of galaxies would be valuable.

The sharp drop in Fe abundance in the outermost region is, given the potential for systematic or statistical uncertainties, difficult to interpret. These include limited photon counts and statistical challenges associated with low-temperature systems, where Fe abundance must be inferred from only L-shell emissions \citep{kim2011metal}. Assuming accurate measurements, one plausible explanation for this drop is that the observed X-rays originate not only from the ISM but also from the less enriched IGrM of the fossil group dominated by Mrk1216. This hypothesis is supported by the fact that Mrk1216’s surface brightness profile is better described as a double-$\beta$ model typical to cool-core systems. However, Mrk1216 lacks a cool core, and both models in the fit are broad, with the second dominating beyond the fourth radial bin ($\sim 103$kpc). This radius coincides with onset of the decline in Fe abundance, suggesting that the second $\beta$ component may represent IGrM emissions. That said, the absence of measurements for other elements severely limits our ability to make a reliable assessment regarding this hypothesis.

\subsection{Relative SNe Contributions}

The abundance ratios observed in Mrk1216 suggest that SNcc contribute more to the total abundances than do SNIa. The \textit{SNeRatio} fit results show the calculated ratios (see Table \ref{tab:SNeRatio_fit_results} and Figure \ref{fig:Mrk1216_sneratio_fits}). The [Mg/Fe] ratio measured in the central two regions exceeds the value predicted by the model. Since a higher [Mg/Fe] ratio is characteristic of a relic galaxy, our ISM abundance measurements align with these expectations.

SNe ratio estimates become more reliable when a wider range of elements are included in calculations. However, observational limitations and instrumental constraints led us to use a small number of ratios—[Mg/Fe], [Si/Fe], [S/Fe]—which increases uncertainty. This challenge is particularly pronounced in low-temperature systems like galaxies, where lower emissivity limits the ability to detect the emission lines used to determine element abundance (i.e., Ar, Ca, Ni) compared to clusters. Next generation satellites equipped with higher-resolution instruments will enable more comprehensive measurements. In the meantime, to make the most of the available data, we also used the spectrum  obtained from the RGS instrument to measure the SNe ratio, which allowed us to include additional elements measurable by RGS but not EPIC (i.e., [O/Fe], [Ne/Fe]). Although the spectral regions differ, the nearly flat radial SNe profile calculated from EPIC spectra allows us to expect similar results from the RGS. Despite being somewhat higher at $0.21_{-0.06}^{+0.09}$, we obtained a consistent SNIa ratio within uncertainties, as was expected.

Apart from this, evaluating the theoretical background of our model suggests that the fit may be improved by adopting a more realistic IMF. We employed a simple and uniform IMF—the Salpeter IMF \citep{salpeter1955luminosity}—to calculate SNcc yields, in order to avoid systematic differences when comparing with other results. However, in the specific case of Mrk1216, it is well established that the stellar population of the galaxy exhibits a bottom-heavy IMF \citep{ferre2017two}. To reflect this, we refitted the abundance ratios in the innermost region using a steeper IMF slope (2.8). This adjustment led to a slight reduction in the quality of the model fit, with the stat./dof value increasing from $11.14/2$ to $11.40/2$ and the SNIa contribution ratio decreasing from 16.7 to 14.2, which did not contribute to resolving the discrepancy between the observed abundance patterns and the model predictions. Conversely, using a top-heavy IMF (slope of 2.0) led to an increase in [Mg/Fe] ($\approx1.3\%$), slightly reducing this discrepancy, with stat./dof = $10.96/2$ and an SNIa ratio of 18.8; still, these changes are too minor to draw any definitive conclusions.

It should be noted that the IMF used for modelling the ISM represents the cumulative stellar population responsible for enriching the hot plasma since its formation, whereas the IMF inferred from stellar analyses reflects the currently surviving stellar populations. Therefore, it is plausible that the IMF derived from the ISM differs from that of the stellar population, as it is more top-heavy in our case. Investigating such IMF variations would require more comprehensive studies with additional elemental abundance measurements, which is beyond the scope of this work.

Finally, canonical IMFs are often known to fall short in reproducing galactic enrichment trends, as they fail to capture the full complexity and variability inherent in the chemical evolution of galaxies \citep{cappellari2012systematic, van2012stellar, martin2018timing, yan2021downsizing, dib2022galaxy, barbosa2021does}. In this regard, the Integrated Galaxy-wide Initial Mass Function (IGIMF) framework provides a more flexible and physically motivated alternative. By allowing the IMF to vary with the star formation rate, the IGIMF offers a more realistic modelling of chemical enrichment and star formation histories \citep{kroupa2003galactic, weidner2005relation, recchi2009chemical, marks2012evidence, jevrabkova2018impact, yan2019star, hosek2019unusual, yan2021downsizing, dib2022galaxy}. These types of methodological advancements in upcoming research may contribute significantly to resolving the ongoing discrepancies between observations and models, while also encompassing different evolutionary pathways that shape chemical enrichment of galaxies.

\subsection{Enrichment Scenarios}

Another noteworthy aspect is the nearly flat radial SNIa ratio profile of Mrk1216, with an average ratio of $\sim0.17$ up to $\sim0.42R_{500}$. These results are consistent with the average profile of groups and ellipticals in \citet{mernier2017radial}. A flat relative SNIa contribution is often regarded as strong evidence for the early-enrichment scenario \citep{yates2017iron, ezer2017uniform, mernier2017radial, erdim2021relative, gastaldello2021metal}. According to this view, a significant portion of enrichment occurred prior to the formation of large systems such as galaxies or galaxy clusters ($z\sim2$) \citep{oppenheimer2006cosmological}. Consequently, both central and outskirt plasmas, despite having evolved under distinct physical conditions, exhibit a similar relative SNe contribution.

Furthermore, the pre-enrichment scenario implies that the gas enriched within proto-groups and proto-cluster galaxies may have been expelled beyond their shallower potential wells at even earlier times (at $z >= 3$) \citep[and references therein]{gastaldello2021metal, mernier2022cycle}. In our case, although the uncertainties increase in the outskirts, the indication of a flat SNIa profile becomes more pronounced because of the lack of efficient mixing mechanisms in Mrk1216 \citep{buote2019extremely}, as mentioned in Section \ref{mrk1216}. Consequently, plasma from the outskirts predominantly represents the most recently accreted matter, which is distinct from plasma accreted earlier \citep{gastaldello2021metal}.

\section{Conclusions}
\label{conclusions}

In this paper, we presented the results of X-ray analyses of the relic galaxy Mrk1216 using \textit{XMM-Newton} observations. We compared our findings with properties of massive ellipticals reported in previous studies and interpreted them in the context of galaxy formation and chemical enrichment scenarios. Our main conclusions are summarised as follows:

\begin{itemize}

    \item The radial temperature profile of Mrk1216 exhibits a negative gradient—a feature generally associated with dynamically disturbed ETGs according to the classification proposed by \citet{kim2020temperature}. Interestingly, the fossil group NGC6482 \citep{khosroshahi2004old}, another outlier in the same sample, also shows a similar negative profile. The presence of such gradients in relaxed fossil systems suggests a possible connection between fossil nature and the development of negative profiles, offering insights into heating–cooling processes and AGN evolution \citep{gaspari2013chaotic, werner2018digging}.

    \item In the two-phase formation model, galaxies initially assemble through gas-rich mergers that drive rapid star formation. $\alpha$-elements such as Mg, produced in massive stars and expelled by SNcc, yield elevated $[\alpha/Fe]$ ratios—signatures of short and intense star formation. In Mrk1216, we measured a significantly high [Mg/Fe] ratio in the X-ray emitting ISM, consistent with earlier stellar abundance results \citep{ferre2017two}. This agreement reinforces the view that Mrk1216 still preserves its red-nugget-like nature and has not transitioned into the second phase.

    \item Abundance ratios in Mrk1216 indicate that SNcc dominated over SNIa in chemical enrichment ($R_{Ia} \approx 0.17 \pm 0.01$). However, the measured [Mg/Fe] in the central regions exceeds model predictions, highlighting the need for further investigation. Our adoption of a simple, uniform IMF (Salpeter) may contribute to this discrepancy. More realistic IMF prescriptions, such as the IGIMF, are recommended for improved chemical evolution modelling.

    \item Mrk1216 exhibits a notably flat radial SNIa ratio profile up to $\sim0.42R_{500}$, consistent with an early-enrichment scenario \citep{werner2013uniform, mantz2017metallicity, biffi2017history}. This suggests that a significant portion of enrichment occurred before system assembly, with relatively similar SNe contributions in both the central and outer plasmas. The lack of efficient mixing mechanisms in Mrk1216 further amplifies the flat profile, implying that the outer plasma may represent the most recently accreted matter.

\end{itemize}

This study demonstrates that the structural and chemical properties of Mrk1216 are consistent with the predictions of the hierarchical formation model. Although conclusions based on a single object limit the strength of interpretations, this study emphasises how future research adopting a similar perspective can yield valuable insights. Identifying comparable systems and conducting in-depth observations will be key. Nevertheless, many unanswered questions may be addressed with the high–resolution capabilities of upcoming missions like Athena X-IFU (X-ray Integral Field Unit) \citep{barret2016athena} or XRISM (X-ray Imaging and Spectroscopy Mission), the successor to ASTO-H/HITOMI \citep{tashiro2022xrism}. Finally, there is a pressing need to refine existing theoretical frameworks by incorporating more realistic IMF models, including the dust depletion effect in chemical modelling, and the resolution of uncertainties related to SNIa progenitors. Certainly, this list is not exhaustive. All advancements in these areas will contribute significantly to our understanding of chemical enrichment, structural formation, and related fields.


\section*{Acknowledgements}
The authors thank the anonymous referees for their constructive comments and suggestions. We would like to express our sincerest gratitude to Dr. François Mernier, Dr. Turgay Çağlar and Fatih Hazar for their support and insightful contributions throughout the course of this research. This work received financial support from the Scientific and Technological Research Council of Turkey (TÜBİTAK), project number 121F263. Lastly, we acknowledge that this paper forms part of M. Kıyami Erdim’s PhD dissertation.

\paragraph{Competing Interests}
`None'

\paragraph{Data Availability Statement}

The \textit{XMM-Newton} raw data used in this article are available to download at the HEASARC Data Archive website ( \url{https://heasarc.gsfc.nasa.gov/docs/archive.html}). The reduced data underlying this article will be shared on reasonable request to the corresponding author.


\bibliographystyle{plainnat}  
\bibliography{pasguide}  

\end{document}